\documentclass[3p,times]{elsarticle}

\usepackage{amssymb}

\begin{document}

\begin{frontmatter}

\title{Astronomical Data Formats: What we have and how we got here}

\author{Jessica D. Mink}

\address{Smithsonian Astrophysical Observatory, 60 Garden St., Cambridge, MA 02131}

\begin{abstract}
Despite almost all being acquired as photons, astronomical data
from different instruments and at different stages in its life
may exist in different formats to serve different purposes.
Beyond the data itself, descriptive information is associated with
it as metadata, either included in the data format or in a larger
multi-format data structure.  Those formats may be used for the
acquisition, processing, exchange, and archiving of data. It has
been useful to use similar formats, or even a single standard to
ease interaction with data in its various stages using familiar tools.
Knowledge of the evolution and advantages of present standards is
useful before we discuss the future of how astronomical data is formatted.
The evolution of the use of world coordinates in FITS is presented as
an example.
\end{abstract}


\end{frontmatter}

\section{Where We Are}
The astronomical community's most widespread data format, FITS
\citep{1981A&AS...44..363W}
is 35 years old, and interest in developing a new and improved standards
for formatting the larger and more varied types of astronomical data
being produced by more and more complicated instruments on larger and
larger telescopes is spreading \citep{2014ASPC..485..351T} and
\citep{Thomas2015}. Several existing options are being proposed: HDF5, a
Hierarchical Data System \citep{Jenness2015}, and JPEG2000, a
widely-used image format \citep{Kitaeff2015}, among others.  The
problems we face are not all new, and I would like to cover some history
about how we got where we are and how our present solutions developed.

\section{Formatting Data Through Its Life-cycle}

\subsection{Genesis: Origins of Astronomical Data}
In the beginning, there is light. Most astronomical observations are of
photons.  In addition to a count or other measure of their intensity,
we record information including direction of the source, wavelength,
polarization, or frequency or energy of individual photons or some
grouping thereof, and the time(s) at which they were collected.
Associated metadata describing the conditions under which the data was
created may be included with the data as a header, trailer, or internal
labels, or may reside in a separate format, such as a logbook, digital
table, or label on the data container.

At its simplest, a data format includes data structured in some way to
make it retrievable.  It may be a qualitative drawing in a logbook, such
as Galileo's drawing of Jupiter and its largest satellites, with descriptive
information about the data written right next to it.  It may be a
table of numbers in a published paper or monograph, with the text of the
paper providing the contextual metadata and the headings on the table
labeling the actual numbers.

In the nineteenth century, it became possible to record a signal from
photons from the sky more directly on glass photographic plates, such as
those in the Harvard Plate Collection \citep{2009ASPC..410..101G}.
It is made up of photographic plates
containing images of the sky, with metadata as notes in logbooks and
on the paper jackets in which the plates are stored. Metadata for each
plate includes the pointing direction, the time and exposure of the
observation, the name of the object being observed, and who observed it.
The logbook and jacket indicate what telescope was used and where it was
located (See Figure 1).  Sky coverage comes from the telescope focal plane
plate scale and the physical size of the plate.  Not all useful parameters
were (or needed to be) written out because the humans using the data knew
them, so additional work has been needed to make the plate images
scientifically usable \citep{2013PASP..125..857T}.

\begin{figure}[h!]
\centering
\includegraphics[scale=0.40]{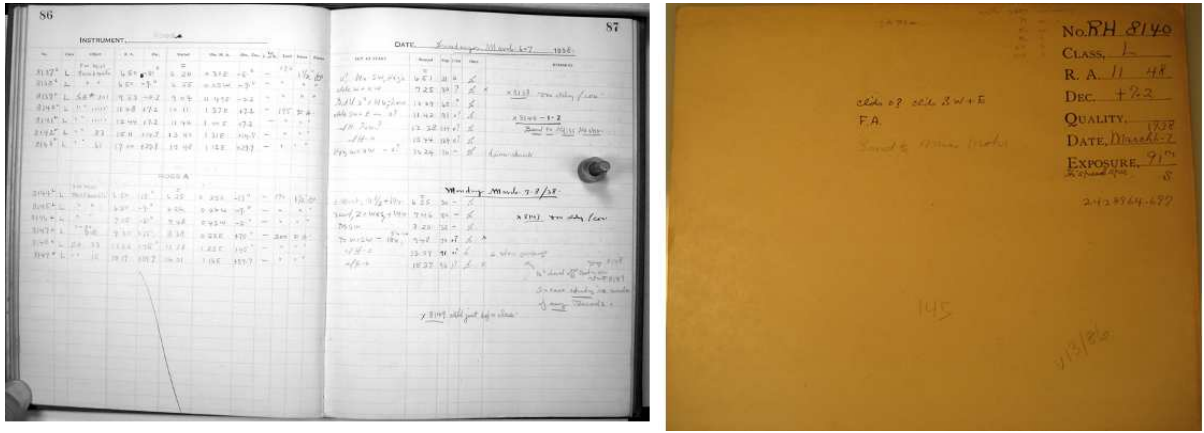}
\caption{Harvard plate metadata in logbook and on plate jacket}
\end{figure}

Digital data acquisition formats are usually limited to one instrument
or a class of instruments. Metadata is usually recorded digitally
either in the same files or files associated in some way within a data
structure so that processing software can learn as much as possible from
its digital input.  For spacecraft observations, this would be digital
telemetry; for optical telescopes, this is likely to be some sort of
image files.  Both might have associated input files,
such as pointing catalogs or fiber positions.

\subsection{Transfer and Exchange}
As we process that data, we end up with derived data which can take
many different forms.  To exchange that data, process it with standard
software, and archive it so scientists can read it in the near or far
future, the metadata has to be standardized and well-documented.  It is
helpful if that metadata travels easily with the actual bits of data.

\subsection{Processing and Analysis}
But the data we have acquired is not ready to be used for science.
As David Hogg noted \citep{Hogg2013}, "The data is like a noisy hash of the
things we care about." It has to be processed from raw data,
including observations and calibrations, to something that can be
analyzed.  Before we can look at the data, we want it formatted in a
specific way which makes it accessible to the tools we wish to use.

In the course of processing, metadata as well as data is changed, with
information about the processing or results of the analysis being
added.  It is useful if the metadata both utilizes
standard definitions and is clearly associated with the data as part of
the same file or data structure.

\subsection{Archiving Data}
Final disposition of data can take several paths. It can be destroyed
because it is seen to have no value or there is no space to keep it.
It can be stored on media which become unreadable, such as no-longer-readable
or slowly-decaying tape or disk formats). It can be presented in a
publication, copies of which are preserved in multiple places, though
only that part of the data content relevant to the publication will be
preserved. And finally, data may be preserved in a format which is both
persistent in content and readability.

Most ground-based and much space-based data from the 1970's
through the 1990's was recorded on magnetic tapes or disks which are now
next to impossible to read.  More recent data on CDROM's and DVD's will
be lost as the media degrade and readers become less common.
While tapes I made during the 1970's and 1980's are no longer readable,
Hollerith cards of photometry and software from my senior year in
college are still human- and scanner-readable (see Figure 2).

\begin{figure}[h!]
\centering
\includegraphics[scale=0.4]{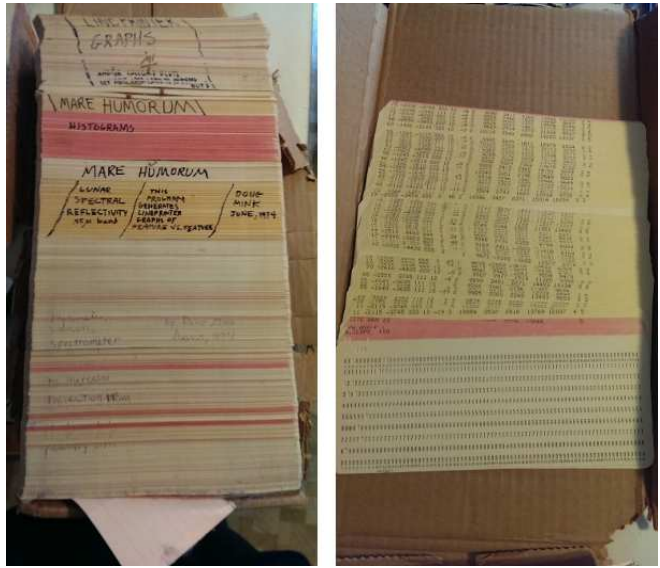}
\caption{The format for the photometry on these cards from 1973 is
documented by the software, also in the card deck, which reads them.}
\end{figure}

If they can, scientists share their data, analysis of it, and
conclusions from it in presentations and papers.  In the age of hardcopy
journals and books, these methods were more persistent than any other
method of preservation, but 21st century astronomers tend to find articles
through ADS \citep{2000A&AS..143...41K} or the ArXiv server
\citep{Vence2014} and read them online instead of in hard copy.
As journals move toward online publication, standardized, retrievable
formatting for the long term is becoming an issue here, too.

We now have the storage capacity to save most of the data
that we take, as well as its derivatives.  So far the most permanent
format we are using is printed paper, with tables and graphs containing data.
Photographic plates, which degrade a bit faster over time, with
metadata in separate paper logs, have lasted over a century. But now most of
our data and more and more of the papers and presentations which
describe its meaning are digital.  We need both persistent media and
persistent formats to keep today's data accessible over time.

\section{Standardizing Data Formats}
As the inventors of FITS noted in their first paper \citep{1980SPIE..264..298G}:
\begin {quote}
Under the traditional system for data interchange in astronomy, each
institution exports data on magnetic tape in its own unique internal
format. Thus, a group of N "cooperating" institutions would begin by
creating N(N-l) format translation programs. Then, whenever one of the
institutions changes its internal format, the other N-1 institutions
have to change their corresponding translation programs. For obvious
reasons, this traditional system has been very inefficient. It would be
very nice if the astronomical community could agree, instead, on a
unique data interchange format.
\end {quote}

It may be in the form of structures, such as a single file with
several FITS extensions \citep{1994A&AS..105...53P}, or a file with
metadata linking to data, such as VO returned data packages
\citep{2004SPIE.5493..262D} containing metadata separate from the
actual data.

FITS (Flexible Image Transport System) \citep{1981A&AS...44..363W} was
originally designed in 1979 as an exchange format, first used by radio
astronomers in the AIPS software system \citep{1999ascl.soft11003A}.
It enabled astronomers to share data without having to maintain separate
translation programs.  At roughly the same time, the more flexible and
more complicated N-Dimensional Data Format (NDF) was being developed in
the U.K.  Tim Jenness has explained the evolution of that system
\citep{Jenness2015a}.

The original simple FITS consisted of a human-readable ASCII header of
80-character lines (matching the width of Hollerith cards then used to
store and use computer software and data) and blocks of binary data
described by the header and system commands. Each of these contained an
integral number of 2880-byte blocks, padded with spaces at the end of
each unit.  A basic set of standard metadata keywords was included with
the original FITS definition.  The use of FITS expanded beyond exchange
and archiving to recording and processing as computers got fast enough
that the time it took to read and write ASCII header information and
convert pixel information into internally-usable bits became increasingly
negligible.

This expanding use was aided by the ability to use FITS reading and
writing libraries such as FITSIO \citep{1991BAAS...23R.936P} and CFITSIO
\citep{1999ASPC..172..487P}
to deal with input and output from local software which worked on the
bits or packages of tools such as AIPS
\citep{1999ascl.soft11003A}, IRAF \citep{1999ascl.soft11002N} (which
started with a propriety format and added FITS
\citep{1996ASPC..101..331Z}),
and WCSTools \citep{2011ascl.soft09015M} which perform sophisticated
operations directly on FITS files.

If we wish to display a FITS file, we can use a variety of tools:
DS9 \citep{2000ascl.soft03002S} for images,
TOPCAT \citep{2011ascl.soft01010T} for tables,
FV \citep{2012ascl.soft05005P} for either images or tables,
and something like WCSTools \citep{2011ascl.soft09015M} IMHEAD to check
out the metadata.

\section{Standardizing Metadata: World Coordinate Systems}
The availability of the FITS data format standard with its standard header
format for metadata and system of registered conventions enabled
astronomers to concentrate on science and using its simple 
\texttt{keyword = value} metadata format to transmit parameters of
their data and models to users.

One movement to standardize metadata has been the inclusion of a set
of parameters linking pixels in an image to pointing directions in the
sky or in the case of spectra, specific wavelengths or energies.
As more precise relationships between image and spatial direction
became necessary for specific projects, a variety of world coordinate
system (WCS) solutions were developed. Because there was a standard way
of defining parameters within a FITS header, it was straightforward to
implement these projections in other software.

The original FITS format \citep{1981A&AS...44..363W} included only a
linear projection defined by the keywords CRPIXn, the coordinate system
reference pixel for axis n, CRVALn, the coordinate system value for axis n
at that reference pixel, CDELTn for the the coordinate increment along axis n, 
CROTAn for the rotation angle of coordinate system axis n (usually only
defined for one axis and assumed to be the same for both), and CTYPEn for
the name of the coordinate axis n.

For the 1983 IRAS satellite \citep{1988iras....1.....B}, a set of standard
projections--GNOMONIC for sky sections, AITOFF for all-sky images, and
SINUSOIDAL for the galactic plane--were used in distributed images, with
parameters and comment lines including Fortran code defining the projection
in the image headers. 

\begin{table}[ht!]
\centering
\begin{tabular}{ l
}
COMMENT   PROJECTION FORMULAE: \\
COMMENT    FORWARD FORMULA; XLON0 IS THE CENTER LONGITUDE OF THE \\
COMMENT        MAP.  ARC-SINE AND ARC-COSINE FUNCTIONS ARE REQUIRED. \\
COMMENT     R2D = 45. / ATAN(1.) \\
COMMENT     PIX = 2. \\
COMMENT     RHO = ACOS( COS(XLAT) * COS((XLON-XLON0)/2.) ) \\
COMMENT     THETA = ASIN( COS(XLAT) * SIN((XLON-XLON0)/2.) / SIN(RHO)) \\
COMMENT     F = 2. * PIX * R2D * SIN(RHO/2.) \\
COMMENT     SAMPLE = -2. * F * SIN(THETA) \\
COMMENT     XLINE = -F * COS(THETA) \\
COMMENT     IF(XLAT .LT. 0.)  XLINE = -XLINE \\
COMMENT   \\
COMMENT    REVERSE FORMULA; XLON0 IS THE CENTER LONGITUDE OF THE MAP. \\
COMMENT          ARC-SINE AND ARC-COSINE FUNCTIONS NEEDED. \\
COMMENT     R2D = 45. / ATAN(1.) \\
COMMENT     PIX = 2. \\
COMMENT     Y = -XLINE / (PIX * 2. * R2D) \\
COMMENT     X = -SAMPLE / (PIX * 2. * R2D) \\
COMMENT     A = SQRT(4.-X*X-4.*Y*Y) \\
COMMENT     XLAT = R2D * ASIN(A*Y) \\
COMMENT     XLON = XLON0 + 2. * R2D * ASIN(A*X/(2.*COS(XLAT))) \\
COMMENT \\
COMMENT   REFERENCES: \\
COMMENT    IRAS SDAS SOFTWARE INTERFACE SPECIFICATION(SIS)  623-94/NO. SF05 \\
COMMENT    ASTRON. ASTROPHYS. SUPPL. SER. 44,(1981) 363-370 (RE:FITS) \\
COMMENT   RECONCILIATION OF FITS PARMS W/ SIS SF05 PARMS: \\
COMMENT    NAXIS1 = (ES - SS + 1); NAXIS2 = (EL - SL + 1); \\
COMMENT    CRPIX1 = (1 - SS);      CRPIX2 = (1 - SL) \\
\end{tabular}
\caption{Aitoff all-sky projection information for converting between
image coordinates PIXEL,LINE and sky coordinates XLON,XLAT from the
header of an IRAS image}
\end{table}

In the AIPS system, a set of "Non-Linear Coordinate Systems"
\citep{AIPS27}--TAN (tangent plane), SIN (sinusoidal,
typically along the galactic equator), ARC, and NCP (centered on
the north celestial pole)--were developed.  By 1986 \citep{AIPS46},
four more were added: GLS (Sanson-Flamsteed sinusoidal), MER (Mercator),
AIT (Hammer-Aitoff equal area all-sky), STG (Stereographic or zenithal
orthomorphic).  Table 1 shows how an AIT projection is presented in a
FITS header.

In 1993 and 1994, the Space Telescope Science Institute released the
"Digitized Sky Survey", which included astrometric solutions
\citep{1990AJ.....99.2059R} in FITS images extracted from compressed
scans of Schmidt plates from Palomar and ESO.  This was the first
attempt to provide an astometrically accurate WCS for astronomical
images, a problem which has become more acute over time.

After much discussion in the community, Marc Calabretta and Eric Greisen
\citep{2002A&A...395.1061G} presented a more completely worked out set
of 25 world coordinate system transformations, along with a code library
which implemented them \citep{2011ascl.soft08003C}.
Figure 3 shows the how well such a tangent plane projection can be fit
matching stars in an image to a catalog using WCSTools.

\begin{figure}[h!]
\centering
\includegraphics[scale=0.25]{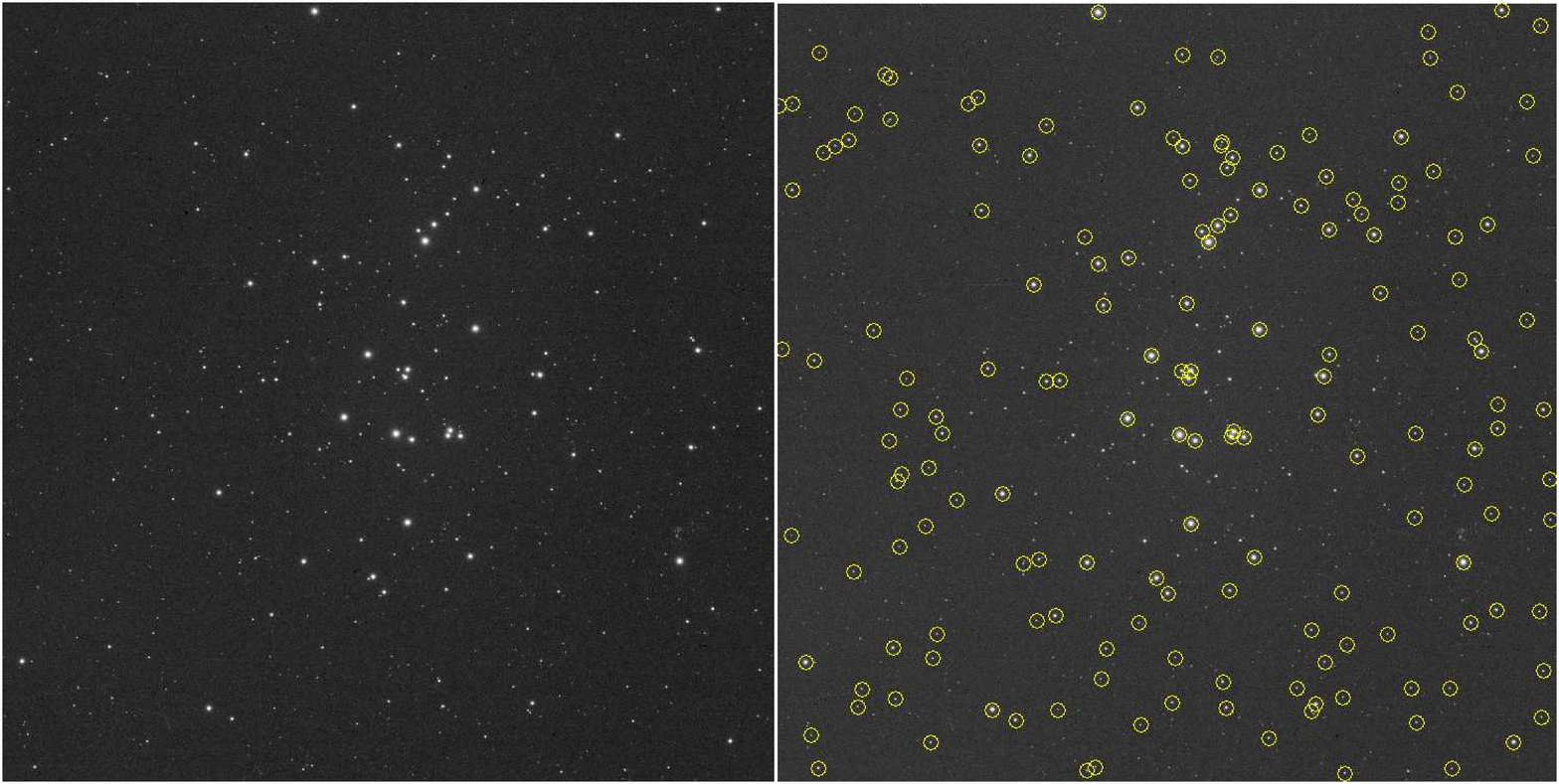}
\caption{The same scanned image of a Harvard photographic plate bare and
with the Tycho 2 catalog overplotted after fitting a FITS WCS TAN
projection using WCSTools}
\end{figure}

But there were still problems exactly matching telescope images, and
several methods of adding information to FITS headers to refine their
astrometry were proposed.  Calabretta, Valdes, and Greisen started in
2003 \citep{2004ASPC..314..551C} with an expansion on the previous FITS
WCS papers, but a resulting standard has yet to be published.

In the mean time, the Spitzer space observatory needed to release data
with improved WCS information and developed the SIP convention
\citep{2005ASPC..347..491S}, while NOAO established the IRAF-understandable
polynomial expansions TNX \citep{TNX2008} (based on FITS-WCS TAN) and ZPX
\citep{ZPX2008} (based on FITS-WCS ZPN).
These registered conventions are documented with the FITS standard at
the FITS Support Office at NASA/Goddard Space Flight Center
\citep{NASAFITS}.  In France, the Terapix project also needed to manage
distortion, and Emanuel Bertin wrote SCAMP \citep{2010ascl.soft10063B} 
to produce a distortion correction to an image's WCS.  WCSTools
includes subroutines to decode all of these distortion methods (see
Table 2), and
the AST library \citep{2014ascl.soft04016B}, IRAF
\citep{1999ascl.soft11002N}, and Astrometry.net
\citep{2012ascl.soft08001L} implement some of them.

\begin{table}[ht!]
\centering
\begin{tabular}{ l l l l
}
\textbf{Code} & \textbf{Projection} & \textbf{Code} & \textbf{Projection} \\
\hline
PIX & Pixel WCS & COO & Conic Orthomorphic \\
LIN & Linear projection & BON & Bonne \\
AZP & Zenithal/Azimuthal Perspective & PCO & Polyconic \\
SZP & Zenithal/Azimuthal Perspective & SFL & Sanson-Flamsteed (Global Sinusoidal) \\
TAN & Gnomonic = Tangent Plane & PAR & Parabolic \\
SIN & Orthographic/synthesis & AIT & Hammer-Aitoff \\
STG & Stereographic & MOL & Mollweide \\
ARC & Zenithal/azimuthal equidistant & CSC & COBE quadrilateralized Spherical Cube \\
ZPN & Zenithal/azimuthal Polynomial & QSC & Quadrilateralized Spherical Cube \\
ZEA & Zenithal/azimuthal Equal Area & TSC & Tangential Spherical Cube \\
AIR & Airy & NCP & Special case of SIN from AIPS \\
CYP & CYlindrical Perspective & GLS & Same as SFL from AIPS \\
CAR & Cartesian & DSS & Digitized Sky Survey plate solution \\
MER & Mercator & PLT & Plate solution (SAO corrections) \\
CEA & Cylindrical Equal Area & TNX & Tangent Plane (NOAO corrections) \\
COP & Conic Perspective & ZPX & Zenithal Polynomial (NOAO corrections) \\
COD & Conic equidistant & TPV & Tangent Plane (SCAMP corrections) \\
COE & Conic Equal area & TAN-SIP & Tangent Plane (Spitzer corrections) \\
\end{tabular}
\caption{WCS projections supported by WCSTools}
\end{table}

\section{Into the Future}
Most of the observational astronomical community uses the FITS format for
at least some stage in the life of their data. FITS meets their needs
by being a well-defined format with embedded human-readable metadata,
and having multiple software packages capable of reading it, published
specifications, and an evolving, well-defined, published set
of metadata models and keywords.  But it is not everything to everybody.
Where do we go from here, and how do we keep the advantages of a format
which has reached across disciplines and uses over decades?  As Brian
Schmidt has said \citep{Schmidt2013}, " Getting standards for data in
place that work requires a consensus dictatorship.  It requires
collaborations between librarians, and computer scientists to figure out
how to create and maintain data hierarchies." As we move forward, we
should be careful not to lose what we have.

\section{Acknowledgements}
Thanks to the Harvard Plate Collection for getting me involved in its
digitization and providing useful images and to Bob Mann and Tim Jenness
for comments which were very useful in developing this paper from a talk at
the 2014 Astronomical Data Analysis Software and Systems conference
which is summarized in \citep{2014arXiv1411.0996M}.

\bibliographystyle{elsarticle-num}
\bibliography{mink}

\end{document}